\documentclass[preprint,12pt]{elsarticle}

\usepackage{amssymb}

\usepackage{url}

\journal{Expert Systems with Applications}

\usepackage[nolist,nohyperlinks]{acronym}
\usepackage{caption}
\usepackage{subcaption}
\usepackage{cleveref}
\usepackage{tabularx}

\begin{acronym}
    \acro{AI}{Artificial Intelligence}
    \acro{RS}{Recommender System}
    \acro{GRS}{Group Recommender System}
    \acro{GR}{Group Recommendation}
    \acro{DL}{Deep Learning}
    \acro{CF}{Collaborative Filtering}
    \acro{MF}{Matrix Factorization}
    \acro{MAE}{Mean Absolute Error}
    \acro{KNN}{K Nearest Neighbors}
    \acro{PMF}{Probabilistic Matrix Factorization}
    \acro{URP}{User Ratings Profile Model}
    \acro{NN}{Neural Network}
    \acro{NCF}{Neural Collaborative Filtering}
    \acro{IPA}{Individual Preference Aggregation}
    \acro{GPA}{Group Preference Aggregation}
    \acro{GMF}{Generalized Matrix Factorization}
    \acro{MLP}{Multi-Layer Perceptron}
    \acro{CNN}{Convolutional Neural Network}
    \acro{RNN}{Recurrent Neural Network}
    \acro{LSTM}{Long Short-Term Memory}
    \acro{GGMF}{Group Generalized Matrix Factorization}
    \acro{GMLP}{Group Multi-Layer Perceptron}
    \acro{ML}{Machine Learning}
    \acro{DLR}{Deep Reinforcement Learning}
    \acro{CARS}{Context-Aware Recommender Systems}
    \acro{CBOW}{Continuous Bag of Words}
\end{acronym}
\acrodefplural{RS}[RSs]{Recommender Systems}

\begin{document}

\begin{frontmatter}



\title{Deep Neural Aggregation for Recommending Items to Group of Users}


\author[upm,knodis]{Jorge Dueñas-Lerín}
\author[upm_si,knodis]{Raúl Lara-Cabrera}
\author[upm_si,knodis]{Fernando Ortega}
\author[upm_si]{Jesús Bobadilla}

\address[upm]{Universidad Politécnica de Madrid, Madrid, Spain}
\address[upm_si]{Departamento de Sistemas Informáticos, Universidad Politécnica de Madrid, Madrid, Spain}       
\address[knodis]{KNODIS Research Group, Universidad Politécnica de Madrid, Madrid, Spain}

\begin{abstract}
Modern society devotes a significant amount of time to digital interaction. Many of our daily actions are carried out through digital means. This has led to the emergence of numerous Artificial Intelligence tools that assist us in various aspects of our lives. One key tool for the digital society is Recommender Systems, intelligent systems that learn from our past actions to propose new ones that align with our interests. Some of these systems have specialized in learning from the behavior of user groups to make recommendations to a group of individuals who want to perform a joint task. In this article, we analyze the current state of Group Recommender Systems and propose two new models that use emerging Deep Learning architectures. The experimental results demonstrate the improvement achieved by employing the proposed models compared to the state-of-the-art models using four different datasets. The source code of the models, as well as that of all the experiments conducted, is available in a public repository.
\end{abstract}




\end{frontmatter}


\section{Introduction}\label{sec:introduction}

Recommendation of products and services is a very relevant issue in our society. Many important tech companies make explicit or implicit recommendations of music (Spotify)~\cite{chen2020hpcf}, travels~\cite{logesh2019efficient} (TripAdvisor), videos (TikTok or YouTube), movies (Netflix), news (Facebook), products (Amazon), etc. For this reason, \acp{RS} are an important field in the \ac{AI} scenario. \Ac{RS} research has traditionally focused on improving accuracy, that is, providing predictions and recommendations as similar to the user's preferences as possible. To achieve this objective, mainly two areas have been addressed: 1) creating improved algorithms and machine learning models, and 2) incorporating as much information as possible: demographic, social, contextual, etc.

Regarding the second area, \ac{RS} can be classified according to their filtering strategies: demographic-based \ac{RS} recommends to the active user those items (products or services) that other users with similar demographic~\cite{yannam2023improving} features (such as sex or age) liked. Content-based filtering~\cite{perez2021content} recommends items whose content (text descriptions, images, tags, etc.) is similar to those consumed by the active user. Social filtering~\cite{sun2020leveraging,song2020deep} relies on the followed, following, friends, etc. lists. Context-based filtering\cite{valera2021context} takes into account a diverse set of information, such as user location, emotions, and physical conditions. The most accurate and popular strategy in \ac{RS} is \ac{CF}~\cite{bobadilla2022neural}, which relies on a dataset containing the preferences of each user regarding the set of available items: songs, movies, books, etc. Note that: a) these preferences can be explicit (votes, number of stars) or implicit (listened songs, watched movies, clicked URL), b) the preferences can be stored in a two-dimensional matrix where each row represents a user, each column represents an item, and each value represents the rating or the absence of rating, and c) the matrix of ratings will be extraordinarily sparse, since users can only consume a little proportion of the available items.

\begin{figure}
    \centering
    \begin{subfigure}[b]{0.45\textwidth}
        \centering
        \includegraphics[width=\textwidth]{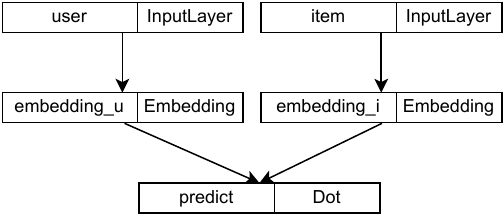}
        \caption{\ac{GMF}}
        \label{fig:gmf}
    \end{subfigure}
    \hfill
    \begin{subfigure}[b]{0.45\textwidth}
        \centering
        \includegraphics[width=\textwidth]{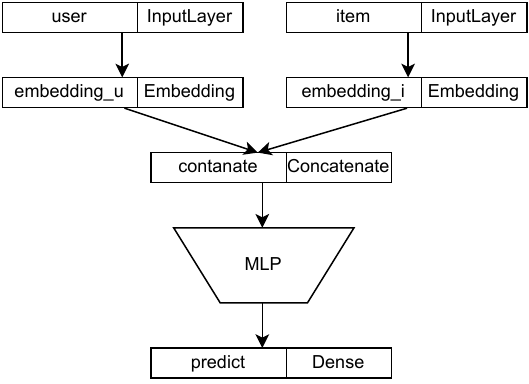}
        \caption{\ac{MLP}}
        \label{fig:mlp}
    \end{subfigure}
    \caption{\ac{NN} architectures for \ac{CF}.}
    \label{fig:dl-based-cf}
\end{figure}

Regarding the algorithms and machine learning models that \ac{RS} use, the \ac{KNN} algorithm was widely chosen to implement the \ac{CF} approach, since its design directly fits the \ac{CF} concept. The main drawbacks of \ac{KNN} are lack of accuracy and poor performance, since it is a memory-based method. The \ac{MF} machine learning model~\cite{bin2021matrix} successfully replaced the \ac{KNN} method; it iterate to learn a representation of users and items in the latent space, coded as hidden factors, then uses them to make predictions by means of a dot product. \ac{MF} has proven to be very accurate and performs fine; their main drawbacks are that as all the model-based approaches it is necessary to update periodically the model, and that the user and item hidden factors combination function: the dot product is linear and it does not completely catch the complex relations existing among users and items. Currently, \ac{MF} is replaced by different \ac{DL} models. \Cref{fig:dl-based-cf} contains a graphical representation of these models. \ac{GMF}~\cite{xue2017deep} model (\cref{fig:gmf}) mimics the \ac{MF} hidden factors by means of an embedding to code users and another embedding to code items, both in the same latent space, as the \ac{MF} does; a dot product is also used to obtain predictions. To improve accuracy, the \ac{MLP} model~\cite{he2017neural} (\cref{fig:mlp}) replaces the dot product with a \ac{MLP} that better combines both embeddings in a nonlinear way.

The most accurate \acp{RS} process \ac{CF} data using some \ac{MF} or \ac{DL} model, and they combine the \ac{CF} approach with some other approaches (social, demographic, content, etc.), depending on the available data in their context. It is normal to join data using an ensemble~\cite{liu2023similarity} of several \ac{DL} models, where \ac{CNN} are mainly used to process content images, \ac{RNN} or \ac{LSTM} to process texts, graph-based models for social information, and \ac{MLP} to integrate demographic data~\cite{huang2022twosteps_graphinfo_hybrid}. In addition to \ac{DL} methods, \ac{DLR} approaches have been explored~\cite{chen2023DRLsurvey} to improve the recommendation in environments where real-time user feedback occurs.

As shown, \ac{RS} have evolved over time, incorporating state-of-the-art models and considering all the available data. As a result, accuracy has improved towards a point that recommendations are positively valued for users, and where it is difficult to significantly increase accuracy by simply improving the model design or complexity. In this scenario, companies focus on getting more and more user and item information as the most effective way to improve \ac{RS} quality results and revenues through advertisements. Of course, this is not the end of research in \ac{RS} accuracy, but it is time to focus on the sometimes `secondary` research in the area: ``\textit{beyond accuracy}'' aims. Popular beyond accuracy objectives are to provide diverse, novel and reliable recommendations, once accuracy is good enough, users appreciate receiving recommendations covering several areas (diverse), not evident (novel)~\cite{mendoza2020evaluating} and providing some degree of certainty (reliability)~\cite{zhu2018assigning}. Moreover, different useful applications of big data have been addressed to cluster users and items\cite{okuda2017}, tools to navigate through \ac{CF} information, as dynamic trees~\cite{hernando2014hierarchical} of related items and/or users~\cite{wang2017hierarchy}, etc. 

In line with the beyond accuracy objectives, there is a field in the \ac{RS} area whose importance grows due to the social networks, where you can find fellows for a variety of actions: share a car, buy cheaper, do a travel, promote your fair cause or simply go to the movie theater. You can send many messages to find people, update your profiles explaining what you are looking for, but it will probably be a huge help if the \ac{RS} does the job. In these examples, \ac{RS} recommends people to a user and is the prelude to a more popular variation of the same idea: recommend items to a group of users~\cite{dePessemier2014}. In this case, once your group of users is set up, explicitly defined in the dataset, categorized through social recommendations\cite{okuda2017}, or just highly temporal with low-cohesion ephemeral groups\cite{cehvarela2022ephemeral}, it is a challenge to find the products or services that will please the whole set of users in the group. Note that it does not fit directly with the traditional \ac{RS} models, since we have several users instead of just one; at first sight, it would be difficult to incorporate additional hidden factors into the \ac{MF} model, and it seems easier to add embeddings to hold the new users when the \ac{MLP} model is used. In all cases, an aggregation approach is necessary to join users, and a policy is convenient\cite{najafian2020agg} to make fair recommendations and to avoid recommendations of items that a majority like, but a minority of the group really dislike~cite. The design, policy, and placement of the aggregation approach in the model are key to the correct behavior, accuracy, fairness, and performance of the proposed group of user models.
This research has a \ac{CF} approach and uses exclusively the information from the user-item interaction matrix, making it versatile for any recommender in which very little information about users or items is available. Once the group representation is generated, it can be combined with context information if available \cite{xu2020}, or with ensembles of different \ac{DL} approaches like \ac{CNN}\cite{huang2022twosteps_graphinfo_hybrid} or text feature extraction~\cite{roy2022Cat}. The \ac{CF} approach \ac{DL} baselines~\cite{sajjadi2021DeepGroup,duenas2023neural} make a heuristic aggregation of user information, the proposed model relies on an innovative \ac{DL} aggregation stage, which constitutes a theoretical improvement that has been validated in practical tests.


The next sections of the paper are structured as follows: \Cref{sec:related-work} focuses on the most relevant and current research in the group recommendation field; \Cref{sec:model} explains the proposed models, which improve a current deep learning baseline~\cite{duenas2023neural} by incorporating a neural aggregation stage; \Cref{sec:experiments} shows the design of the experiments and explains the results obtained; and \Cref{sec:conclusions} presents the conclusions and future works.

\section{Related work}\label{sec:related-work}

As mentioned above, the design and performance of \ac{GRS} is deeply dependent on three factors: the design of the aggregation approach, its placement in the model, and the aggregation policy. Although the first two factors depend on the model, the aggregation policy is conceptual. When a recommendation for a \ac{GRS} is required, we should determine how to distribute satisfaction among the users of the group~\cite{dara2020survey}. We could maximize the average accuracy (aggregated voting policy), but deviation from the mean is also important, and we probably do not want to highly satisfy the majority at the expense of harming the minority (least misery policy). Recently, a particular weighting of ``\textit{leader}'' users has been proposed~\cite{nozari2020novel}, and these types of user are found using trust information and fuzzy clustering; a very similar approach focuses on genre preferences and clustering, where a relevant contribution is to reduce the cost of clustering~\cite{seo2021group}. Similarly, the average policy can be used as a base, weighted with trusted social networks to correct preferences~\cite{wang2020group}. Fuzzy approaches are also used to combine trust and reputation~\cite{bedi2020combining}. Consensus is a traditional tool to implement policies~\cite{castro2015consensus,meena2018group}. 

The impact of aggregation placement on accuracy and performance was tested~\cite{ortega2013incorporating}, finding that the earlier the aggregation stage is placed, the better the \ac{RS} performance, while the accuracy does not change significantly. \cite{ortega2013incorporating} paper provides memory-based results, and there is no similar paper testing it on model-based approaches, except for \cite{ceh2022performance} where a wide comparative of \ac{GRS} has shown that the average aggregation function and the \ac{MLP} model provide the best accuracy of the combinations tested. Acting on rankings of recommendations is basically to reorder the recommendation lists. Previous work merges (aggregates) the individual rankings of the users in the group to generate a common ranking. Today, this approach is also used, but improved in the model stage; in \cite{pujahari2020aggregation} the \ac{MF} technique based on the preference relation is used to obtain the predicted preference and then combines those preferences using graph aggregation; results report ranking improvements. A genetic algorithm approach is proposed for the \ac{GRS} rankings\cite{meena2018genetic}, where different distances are used as a fitness function. Recommendations can be highly influenced by few members of the group; To avoid it, preference relations are used and improved by a \ac{MF} process to find unknown relations\cite{pujahari2021preference} and then improve \ac{GRS} results.

\Ac{DL} models are very flexible in integrating and aggregating diverse data sources, making them on different layers and thus managing several semantic levels: the deeper the aggregation layer, the higher its dimensionality reduction and then its semantic level. For this reason, it is very different to merge user profiles or item information before they are processed by the model that merge their embedding representations. And it is also different to merge embeddings in the first layers of the model from doing it on its last layers. It is sensible to merge embeddings in a range of the model layers where the information already has been compressed (it provides semantic value), but not too much (it has lost relevant information to make the users or items merge). \ac{GRS} papers focused on the \ac{DL} model's architecture vary their proposals following the explained balance.

\begin{figure}
    \centering
    \begin{subfigure}[b]{0.45\textwidth}
        \centering
        \includegraphics[width=\textwidth]{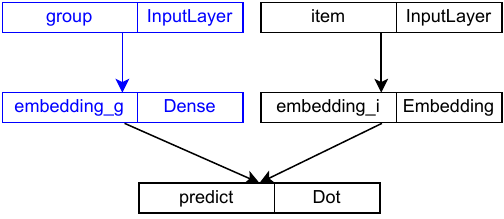}
        \caption{Modified \ac{GMF} model.}
        \label{fig:mo-gmf}
    \end{subfigure}
    \hfill
    \begin{subfigure}[b]{0.45\textwidth}
        \centering
        \includegraphics[width=\textwidth]{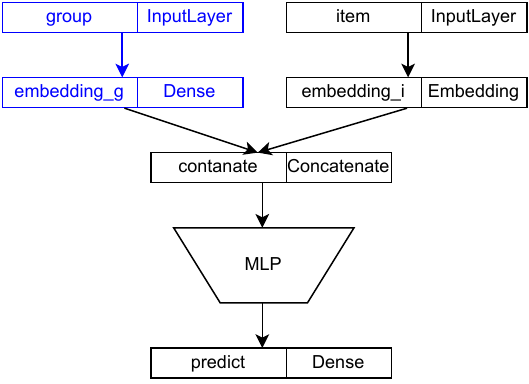}
        \caption{Modified \ac{MLP} model.}
        \label{fig:mo-mlp}
    \end{subfigure}
    \caption{\ac{NN} architectures for \ac{GRS} proposed by~\cite{duenas2023neural}. Blue parts indicates changes with respect to the architectures for single user rating prediction.}
    \label{fig:mo-grs}
\end{figure}

Previous research has identified multiple key points at which the aggregation of user preferences can be carried out~\cite{sajjadi2021DeepGroup} and how to represent it in the latent space~\cite{duenas2023neural}. These studies propose using different embedding layers: one for the items and the other for the group of users. To include these layers, the \ac{GMF} and \ac{MLP} models are slightly modified. \Cref{fig:mo-grs} shows the proposed architectures highlighting in blue the main modifications with respect to their reference models (see \cref{fig:dl-based-cf}). As can be seen, the embedding of the items has not undergone any modification. However, the embedding of users has been replaced by a group embedding. This embedding has been encoded using a dense layer that allows input with multiple active values. The paper proposes different aggregation functions for encoding the input of the network for a group of users. The rest of the network remains unchanged for both the \ac{GMF} model and the \ac{MLP} model. It is important to note that these models are trained with individual ratings and not group ratings, so this previous work proposes how to modify the network input to make predictions for both individual users and groups of users.

A similar approach~\cite{yannam2023improving} provides two \ac{DL} paths: the first processes the data and the second demographic meta-data. Relevant information flows through the first path, where an embedding is set for items, and another embedding is set for groups. It should be noted that the design~\cite{yannam2023improving} does not explicitly feed the network with users belonging to each group. Unfortunately, this baseline is inadequate since its design does not allow the incorporation of new groups in the \ac{RS} without a relearning operation; Our proposed approach feeds the model with each user in the group; consequently, the model learns about users, and it can generalize and predict to new groups without the need to continuously relearn specific patterns. 

\section{Proposed model}\label{sec:model}

The previous work analyzed shows a lot of immaturity in the development of \ac{DL} based \ac{GRS}. Thus, all proposed solutions consist of defining some aggregation mechanism to unify a group of users and represent them as an individual user. For example, \cite{duenas2023neural} defines a function $f(G) \rightarrow \mathbb{R}^{U}$ from which a group of users $G$ is encoded into a $U$-dimensional vector (where $U$ is the number of users in the system), using multihot encoding. In other words, instead of representing a user with a one-hot encoding, as is common when using embeddings, a multihot encoding is employed to represent the group. The main drawback of this model is that all proposed functions $f(G) \rightarrow \mathbb{R}^{U}$ are imposed and are not learned from the data.

\begin{figure}
    \centering
    \begin{subfigure}[b]{0.45\textwidth}
        \centering
        \includegraphics[width=\textwidth]{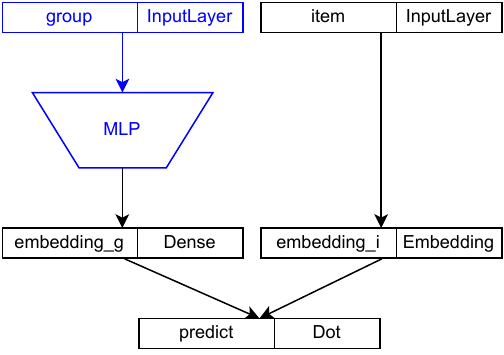}
        \caption{\ac{GGMF}.}
        \label{fig:ggmf}
    \end{subfigure}
    \hfill
    \begin{subfigure}[b]{0.45\textwidth}
        \centering
        \includegraphics[width=\textwidth]{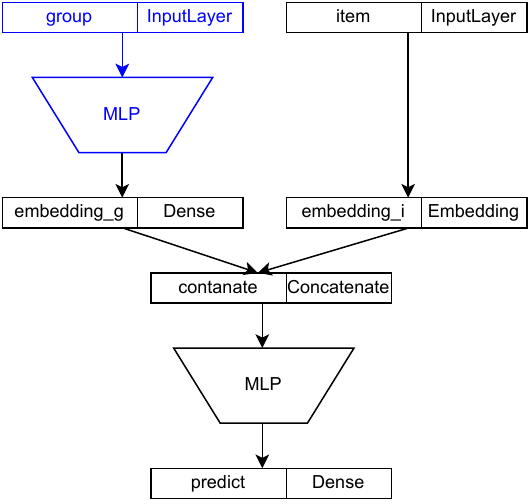}
        \caption{\ac{GMLP}.}
        \label{fig:gmlp}
    \end{subfigure}
    \caption{Proposed \ac{NN} architectures for \ac{GRS}. Blue parts indicates changes with respect to the architectures proposed by~\cite{duenas2023neural}}
    \label{fig:proposed-models}
\end{figure}

The models proposed here go a step further by allowing the aggregation mechanism to be determined by a \ac{ML} model rather than being manually defined. \Cref{fig:proposed-models} presents the proposed \ac{NN} architectures, highlighting in blue changes with respect to the model proposed in~\cite{duenas2023neural} (see \cref{fig:mo-grs}). As can be observed, the \ac{GMF} and \ac{MLP} architectures have been modified to include a \ac{MLP} between the user embedding and the dot product or the concatenate layer, respectively. This \ac{MLP} replaces the function used in~\cite{duenas2023neural} and it allows the \ac{NN} itself adjusts the user aggregation function, deciding the weighting that their votes have on the recommendation that is ultimately made to the group. Technically speaking, this new \ac{MLP} transforms the group of users, represented by a multihot encoding vector, i.e. a vector in which the users than belong to the group are encoded with 1 and the rest of the users are encoded with 0, into the latent space of the users. By connecting it to the dot layer or the concatenate layer of the \ac{GMF} or \ac{MLP} models, respectively, predictions can be generated for the group of users. We name these new models \ac{GGMF} and \ac{GMLP}.

One of the most remarkable features of the proposed \ac{GRS} is that they are capable of performing individual and group predictions. This is achieved by training the individual \ac{GMF} and \ac{MLP} models using individual votes, and then freezing the parameters of these models to train the new \ac{MLP} with group votes. In this vein, to generate individual predictions, we will use the model without considering the new \ac{MLP}. That is, the input to the model will be the user and item to the embedding layer. In contrast, to generate predictions for the group, we will use the same input for the items and encode the multihot representation of the group so that the \ac{MLP} generates its embedding.

\section{Experimental evaluation}\label{sec:experiments}

This section presents the experiments conducted to compare and validate the proposed models. As stated above, the experiments were carried out using two most popular \ac{CF} based \ac{NN} architectures, i.e. \ac{GMF} and \ac{MLP}. These architectures were selected because they are well known and provide the best results for individual predictions. However, the proposed aggregation strategies can be applied to any \ac{NN} architecture that is based on embedding layers.

\subsection{Experimental set-up}

To select the datasets, we considered two factors: a) there are no open datasets available that contain information on group voting; and b) the \ac{GMF} and \ac{MLP} models should be trained using individual voting because the proposed aggregations allow groups predictions to be computed on models already trained for individuals. Therefore, we chose four gold standard datasets in the field of \ac{RS}: \texttt{MovieLens100K} and \texttt{MovieLens1M}, which are the most popular datasets in the field of \ac{RS}; \texttt{FilmTrust}, which is smaller than \texttt{MovieLens100K} and allows us to measure the performance of the aggregation in datasets with a lower number of users, items, and ratings; and \texttt{MyAnimeList}, which has a wider range of votes than \texttt{MovieLens1M}. We did not select other popular datasets such as \texttt{Netflix Prize} or \texttt{MovieLens10M} due to the high computational time required to train and test the models. The train and test splits of these datasets included in CF4J~\cite{ortega2021cf4j,ortega2018cf4j} were used. The main parameters of these datasets can be found in \cref{tab:datasets}.

\begin{table}[ht]
\centering
\scriptsize
\begin{tabularx}{\textwidth}{|l|X|X|X|X|X|}
    \hline
    Dataset        & Number of users & Number of items & Number of ratings & Number of test ratings & Scores     \\ \hline
    FilmTrust      & 1,508           & 2,071           & 32,675            & 2,819                  & 0.5 to 4.0 \\ \hline
    MovieLens100K  & 943             & 1,682           & 92,026            & 7,974                  & 1 to 5     \\ \hline
    MovieLens1M    & 6,040           & 3,706           & 911,031           & 89,178                 & 1 to 5     \\ \hline
    MyAnimeList    & 69,600          & 9,927           & 5,788,207         & 549,027                & 1 to 10    \\ \hline
\end{tabularx}
\caption{Main parameters of the datasets used in the experiments sorted by number of ratings.}
\label{tab:datasets}
\end{table}

The user groups with which the models have been proven have been synthetically generated. To measure the system's performance in different scenarios, groups ranging from 2 to 10 users have been generated. The groups were randomly generated from the rating set. From this rating set, a random item was selected first, and then $G$ users from this set, who had voted for that item, were selected (where $G$ is the group size). \Cref{tab:number-of-groups} contains the number of group ratings generated for model training and test in each dataset and each group size.

\begin{table}[ht]
\centering
\scriptsize
\begin{tabularx}{\textwidth}{|c|X|X|X|X|}
\hline
\begin{tabular}[c]{@{}l@{}}Group\ size\end{tabular} & FilmTrust & MovieLens100K & MovieLens1M & MyAnimeList\\ \hline
2 & 3,948/1,128   & 11,165/3,190   & 124,850/35,672   & 768,639/219,612   \\ \hline
3 & 5,921/1,692   & 16,746/4,786   & 187,275/53,508   & 1,152,958/329,417   \\ \hline
4 & 7,894/2,256   & 22,328/6,380   & 249,699/71,344   & 1,537,277/439,222   \\ \hline
5 & 9,867/2,820   & 27,910/7,975   & 312,124/89,179   & 1,921,595/549,028   \\ \hline
6 & 11,841/3,384   & 33,492/9,570   & 374,549/107,014   & 2,305,914/658,834   \\ \hline
7 & 13,814/3,948   & 39,074/11,164   & 436,973/124,850   & 2,690,233/768,639   \\ \hline
8 & 15,787/4,512   & 44,655/12,760   & 499,398/142,686   & 3,074,552/878,444   \\ \hline
9 & 17,761/5,075   & 50,237/14,354   & 561,822/160,522   & 3,458,871/988,250   \\ \hline
10 & 19,734/5,639   & 55,819/15,949   & 624,247/178,357   & 3,843,190/1,098,055   \\ \hline
\end{tabularx}
\caption{Number of synthetic groups generated for system training and evaluation (train/test) in each dataset. Note that a group is only evaluated for one item.}
\label{tab:number-of-groups}
\end{table}

The training of the models has been carried out as follows. First, individual models (\ac{GMF} and \ac{MLP}) are trained without considering user groups. \Ac{MAE} was used as loss function and Adam as optimizer with a learning rate of $0.001$. A batch size of $64$ was used. There was no epoch limit, but an early stop condition.  Next, the new \ac{MLP} is incorporated into the network. The depth of the \ac{MLP} varies depending on the dataset used. In general, larger datasets require more parameters to be added to the \ac{MLP}. Specifically, a \ac{MLP} consisting of four dense layers ($[64, 32, 16, 8]$) is used for the FilmTrust and MovieLens100K datasets; Since MovieLens1M and MyAnimeList are an order of magnitude larger, they are more complex and have more possible users to be part of the groups, an extra layer with 128 neurons is used ($[128, 64, 32, 16, 8]$); All neurons use ReLU function as activation but in the last layer, which has a linear activation allowing the group representation generated by the \ac{MLP} to produce no restricted latent factors. Finally, the network is trained by updating only the parameters of the new group's \ac{MLP}. In other words, the parameters of the individual models previously trained are not modified during this training. For this training, the following loss function was used:

$$
loss = \frac{1}{\#\Omega} \sum_{<G,i> \in \Omega} | h(G,i) - \hat{r}_{G,i} |,
$$

where $\Omega$ is the set of pairs $<G,i>$ with the ratings of the synthetic groups $G$ to the item $i$, $\hat{r}_{G,i}$ is the prediction of the rating, and $h(G,i) \rightarrow \mathbb{R}$ is a function that unifies the votes of group $G$ for item $i$. Note that the training of the new \ac{MLP} has been carried out independently for each group size.

Finally, the evaluation of the model is performed by measuring the mean squared difference between its predictions and the votes of the users in the groups as follows.

$$
MSE_{G,i} = \frac{1}{\#G} \sum_{u \in G} \left( \hat{r}_{G,i} - r_{u,i} \right)^2, 
$$

$$
MSE = \frac{1}{\#\Omega} \sum_{<G,i> \in \Omega} MSE_{G,i}
$$
    
where $\Omega$ is the set of pairs $<G,i>$ with the ratings of the synthetic groups $G$ to the item $i$, $r_{u,i}$ is the user rating $u$ to the item $i$, and $\hat{r}_{G,i}$ is the prediction of the rating.

\subsection{Experimental results}

\begin{figure}
    \centering
    \begin{subfigure}[b]{1.0\textwidth}
        \centering
        \includegraphics[width=\textwidth]{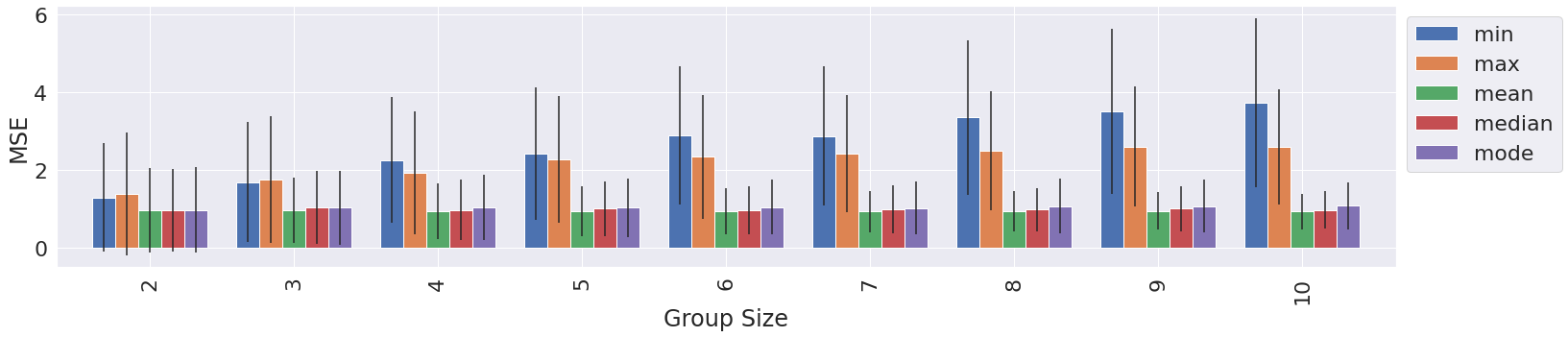}
        \captionsetup{margin={0cm,1.2cm}}
        \caption{MovieLens100K}
        \label{fig:gmf:agg:ml100k}
    \end{subfigure}
    
    \vspace{0.1cm}

    \begin{subfigure}[b]{1\textwidth}
        \centering
        \includegraphics[width=\textwidth]{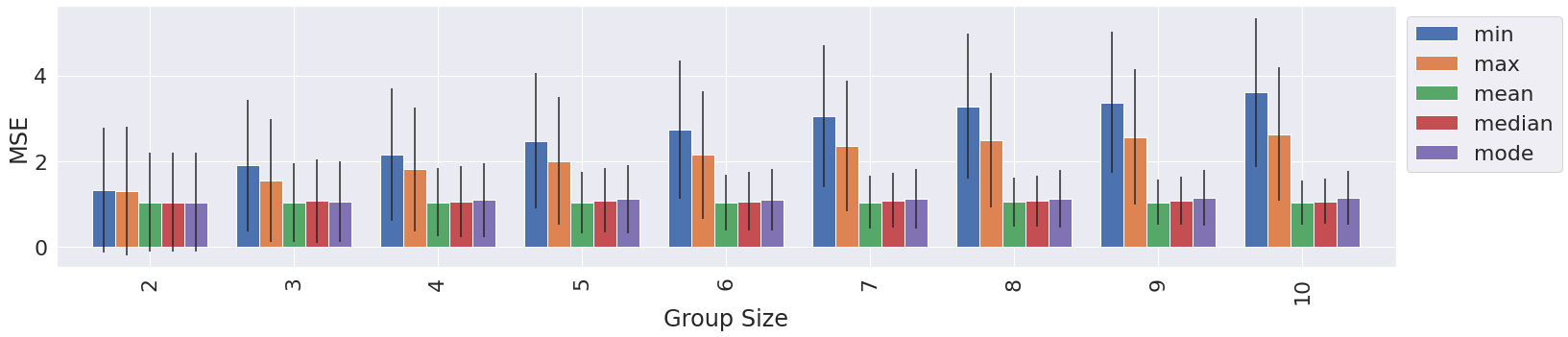}
        \captionsetup{margin={0cm,1.2cm}}
        \caption{MovieLens1M}
        \label{fig:gmf:agg:ml1m}
    \end{subfigure}

    \vspace{0.1cm}

    \begin{subfigure}[b]{1\textwidth}
        \centering
        \includegraphics[width=\textwidth]{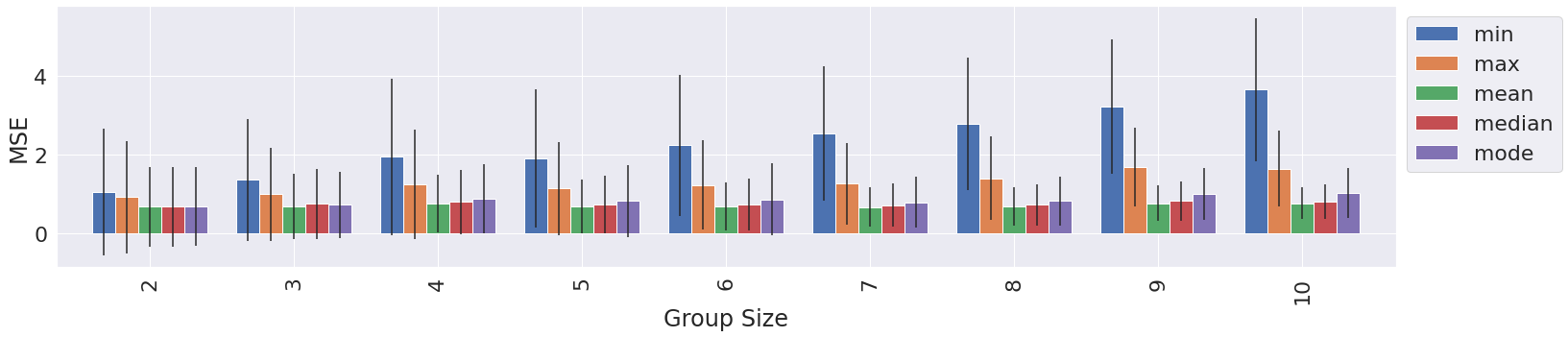}
        \captionsetup{margin={0cm,1.2cm}}
        \caption{FilmTrust}
        \label{fig:gmf:agg:ft}
    \end{subfigure}

    \vspace{0.1cm}
    
    \begin{subfigure}[b]{1\textwidth}
        \centering
        \includegraphics[width=\textwidth]{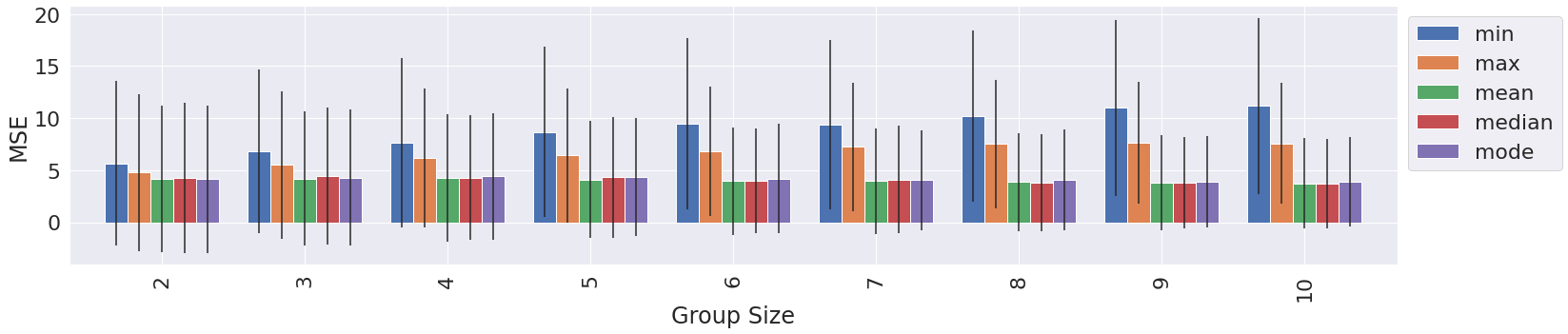}
        \captionsetup{margin={0cm,1.2cm}}
        \caption{MyAnimeList}
        \label{fig:gmf:agg:anime}
    \end{subfigure}
    \caption{Comparison of the error obtained in the test using different functions $h$ for the \ac{GGMF} model.}
    \label{fig:gmf:agg}
\end{figure}

\begin{figure}
    \centering
    \begin{subfigure}[b]{1.0\textwidth}
        \centering
        \includegraphics[width=\textwidth]{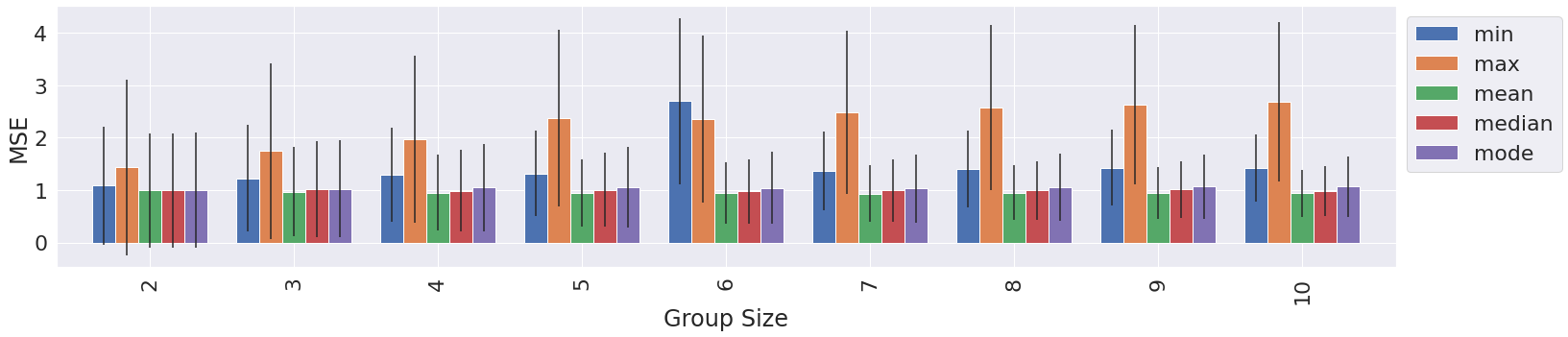}
        \captionsetup{margin={0cm,1.2cm}}
        \caption{MovieLens100K}
        \label{fig:mlp:agg:ml100k}
    \end{subfigure}
    \hfill
    \begin{subfigure}[b]{1\textwidth}
        \centering
        \includegraphics[width=\textwidth]{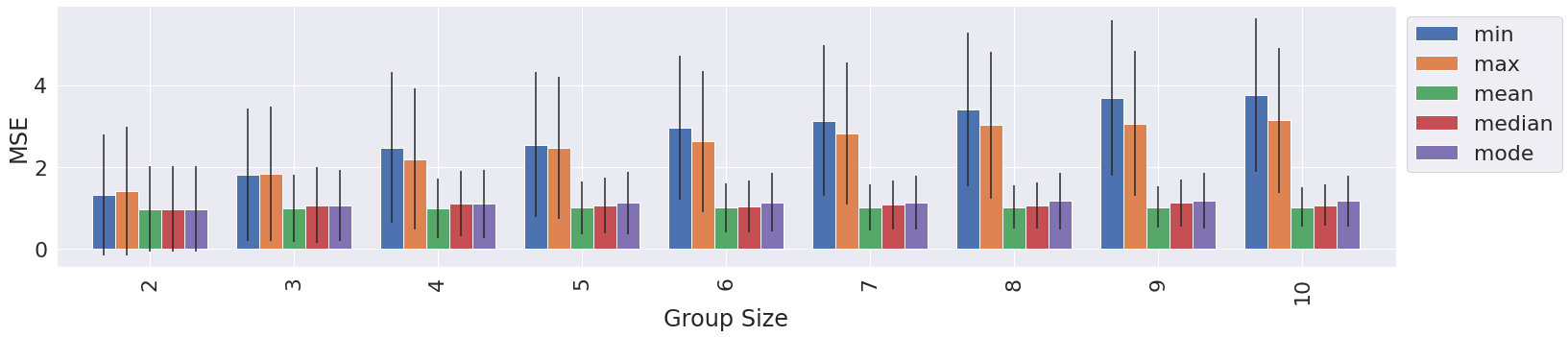}
        \captionsetup{margin={0cm,1.2cm}}
        \caption{MovieLens1M}
        \label{fig:mlp:agg:ml1m}
    \end{subfigure}
    \hfill
    \begin{subfigure}[b]{1\textwidth}
        \centering
        \includegraphics[width=\textwidth]{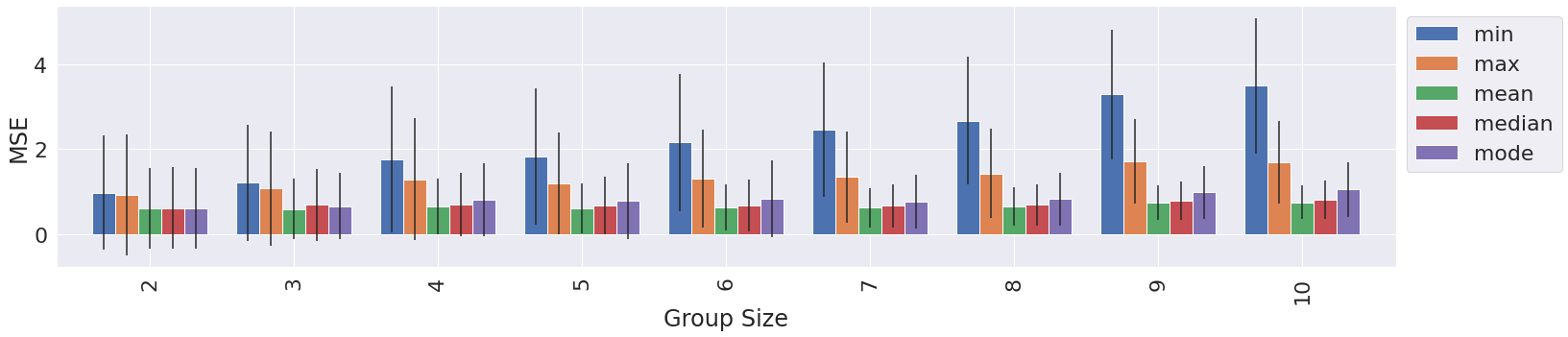}
        \captionsetup{margin={0cm,1.2cm}}
        \caption{FilmTrust}
        \label{fig:mlp:agg:ft}
    \end{subfigure}
    \hfill
    \begin{subfigure}[b]{1\textwidth}
        \centering
        \includegraphics[width=\textwidth]{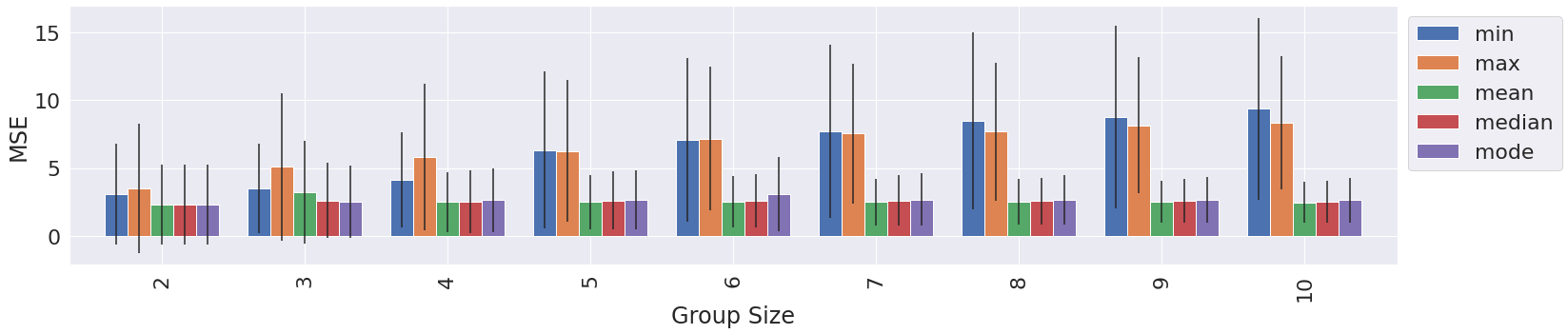}
        \captionsetup{margin={0cm,1.2cm}}
        \caption{MyAnimeList}
        \label{fig:mlp:agg:anime}
    \end{subfigure}
    \caption{Comparison of the error obtained in the test using different functions $h$ for the \ac{GMLP} model.}
    \label{fig:mlp:agg}
\end{figure}

The first experiment conducted aimed to determine the best function  $h(G,i) \rightarrow \mathbb{R}$ to unify the group votes in the loss function during model training. The following functions were tested: minimum group vote (min), maximum group vote (max), mean group vote (mean), median group vote (median) and mode of votes (mode). \Cref{fig:gmf:agg,fig:mlp:agg} display a boxplot with the results of this experiment using the \ac{GGMF} and \ac{GMLP} models, respectively. In all cases, the function that provides the best results (i.e. the lowest error) is the mean, so we fixed $h(G,i) = \frac{1}{\#G} \sum_{u \in G} r_{u,i}$ for the next experiments.

\begin{figure}
    \centering
    \begin{subfigure}[b]{1.0\textwidth}
        \centering
        \includegraphics[width=\textwidth,height=0.25\textwidth]{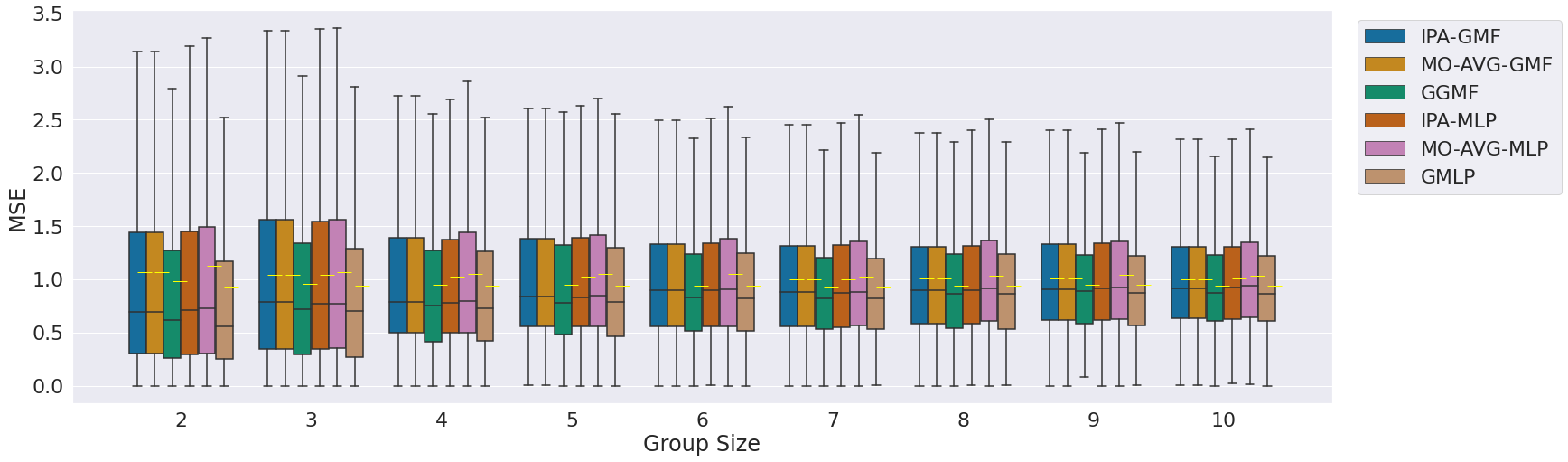}
        \captionsetup{margin={0cm,1.5cm}}
        \caption{MovieLens100K}
        \label{fig:results:ml100k}
    \end{subfigure}

    \vspace{0.1cm}
    
    \begin{subfigure}[b]{1\textwidth}
        \centering
        \includegraphics[width=\textwidth,height=0.25\textwidth]{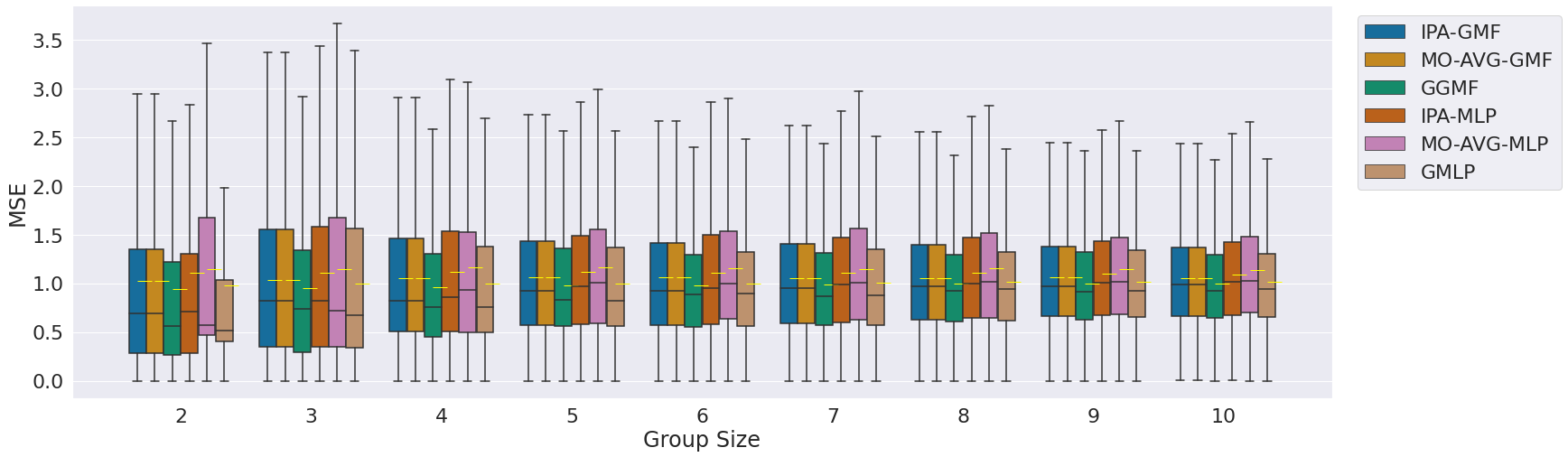}
        \captionsetup{margin={0cm,1.5cm}}
        \caption{MovieLens1M}
        \label{fig:results:ml1m}
    \end{subfigure}

    \vspace{0.1cm}

    \begin{subfigure}[b]{1\textwidth}
        \centering
        \includegraphics[width=\textwidth,height=0.25\textwidth]{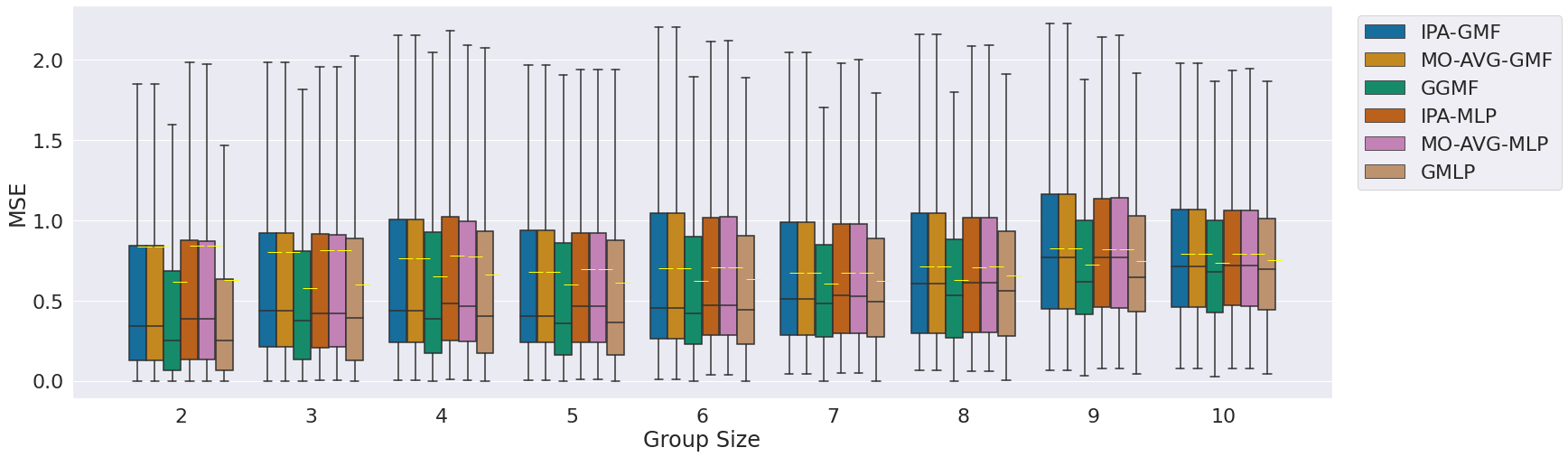}
        \captionsetup{margin={0cm,1.5cm}}
        \caption{FilmTrust}
        \label{fig:results:ft}
    \end{subfigure}

    \vspace{0.1cm}
    
    \begin{subfigure}[b]{1\textwidth}
        \centering
        \includegraphics[width=\textwidth,height=0.25\textwidth]{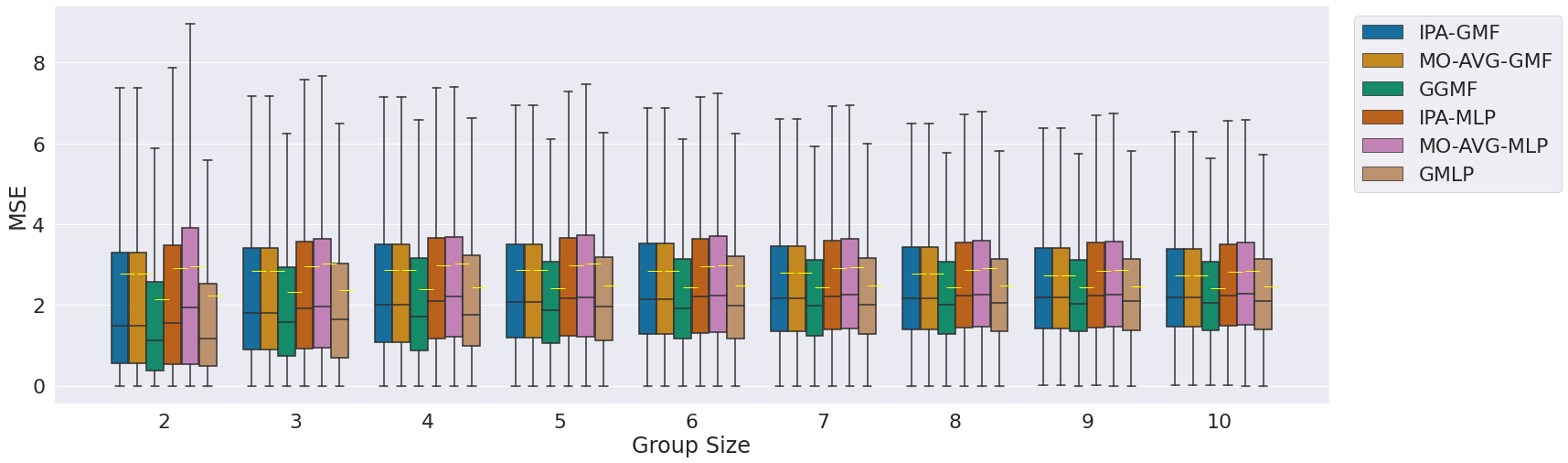}
        \captionsetup{margin={0cm,1.5cm}}
        \caption{MyAnimeList}
        \label{fig:results:anime}
    \end{subfigure}
    \caption{Prediction quality comparison of the proposed models (GGMF and GMLP) with respect to the state of the art baselines.}
    \label{fig:results}
\end{figure}

The following experiment consisted of comparing the performance of the proposed model with respect to other state-of-the-art models. The following baselines were used: the average of individual predictions using the \ac{GMF} model (\texttt{IPA-GMF}), the average of individual predictions using the \ac{MLP} model (\texttt{IPA-MLP}), the multi-hot aggregation function combined with the average proposed by~\cite{duenas2023neural} using the \ac{GMF} model (\texttt{MO-AVG-GMF}), and the same aggregation function with the \ac{MLP} model (\texttt{MO-AVG-MLP}). \Cref{fig:results} presents the results of this experiment for the 4 datasets and different group sizes.

It can be observed that, in general terms, the new group representation made by \ac{MLP} added to the first phase of the model improves the quality of the predictions regardless of the dataset, group size, and predictive individual model used. Therefore, this group representations learned from the user-item interactions over-perform any predefined function selected in advance. The deep group aggregation phase allows us to learn the peculiarities of users, items, and their votes when they act as a group.

All these experiments were conducted using an NVIDIA Quadro RTX 8000 GPU, which boasts 48 GB GDDR6 memory, 4,608 NVIDIA Tensor Cores, and a performance of 16.3 TFLOPS. To ensure reproducibility, the source code, parameter values, and random seeds of all experiments have been made publicly available on GitHub\footnote{\url{https://github.com/KNODIS-Research-Group/dl-cf-groups-deep-aggregation}}.

\section{Conclusions and future work}\label{sec:conclusions}

This work has proposed a new approach to \ac{DL}-based \ac{GRS}. The study suggests extending the most popular \ac{DL}-based \ac{CF} models by adding a \ac{MLP} that allows us to project a group of users into the latent space of users. This projection is then combined with the latent space of the items to obtain predictions. This proposal has been evaluated using two different recommendation models, \ac{GMF} and \ac{MLP}, four different datasets (FilmTrust, MovieLens100K, MovieLens1M, and MyAnimeList), and nine group sizes. All experimental results demonstrate a significant improvement of the proposed models compared to their baselines. Therefore, it can be confirmed that using an \ac{MLP} to learn how to aggregate user groups from the data improves the quality of the predictions.

Furthermore, \ac{GRS} face the challenge of needing more powerful models as the complexity of the dataset increases. To handle complex and diverse data, sophisticated architectures are necessary to capture the nuances and intricacies of user group behavior. On the other hand, the quality of \ac{GRS} remains consistent regardless of the size of the group. If a system demonstrates effective performance for small groups, it can be reliably expected to deliver satisfactory results for larger groups as well, and vice versa. This characteristic highlights the robustness and scalability of such systems, allowing them to cater to various group sizes with consistent recommendation quality.

In future work, it is proposed to redesign the architecture to enable predictions for groups of variable sizes. The proposed models have been trained for a fixed group size, and therefore their results on groups of variable sizes are uncertain. Another line of future work is to incorporate additional information into the datasets, such as friendships or followers, and to assess whether social relationships help or hinder the learning of group preferences.

\section*{Acknowledgments}

This work was partially supported by \textit{Ministerio de Ciencia e Innovación} of Spain under the project PID2019-106493RB-I00 (DL-CEMG) and the \textit{Comunidad de Madrid} under \textit{Convenio Plurianual} with the Universidad Politécnica de Madrid in the actuation line of \textit{Programa de Excelencia para el Profesorado Universitario}.

 \bibliographystyle{elsarticle-num} 
 \bibliography{cas-refs}





\end{document}